\documentclass{aa}
\usepackage{graphicx}
\usepackage{txfonts}
\usepackage{lscape}
\usepackage{natbib}
\bibpunct{(}{)}{;}{a}{}{,} % to follow the A&A style

\begin{document}
   \title{Properties of young star cluster systems: the age signature from near-infrared integrated colours}

   \author{J. F. C. Santos Jr.\inst{1}, H. Dottori\inst{2},
          \and
          P. Grosb{\o}l \inst{3}
          }

   \institute{Instituto de Ci\^encias Exatas, Univ. Federal de Minas
     Gerais, Av. Ant\^onio Carlos 6627, 31270-901 Belo Horizonte, MG, Brazil\\
              \email{jsantos@fisica.ufmg.br}
         \and
             Instituto de F\'isica, Univ. Federal do Rio Grande do Sul,
             Av. Bento Gon\c calves 9500, 91501-970 Porto Alegre, RS, Brazil\\
             \email{dottori@ufrgs.br}
         \and
             European Southern Observatory, Karl-Scharzschild-Str. 2, 85748
             Garching, Germany\\
             \email{pgrosbol@eso.org}
             }

\authorrunning{Santos Jr. et al.}
\titlerunning{Star cluster systems}

   \date{Received 30 October 2012 / Accepted 29 March 2013}

% \abstract{}{}{}{}{} 
% 5 {} token are mandatory
 
  \abstract
   {A recent $JHK_{\rm s}$ study of several grand-design spiral
galaxies, including NGC\,2997, shows a bimodal
distribution of their system of star clusters and star forming complexes
in colour-magnitude and colour-colour diagrams. 
In a comparison with stellar population models including gas, the 
 ($J-H$)$\ vs\ $($H-K_{\rm s}$) diagram reveals that embedded clusters, 
still immersed in their parental clouds of gas and dust, generally have 
a redder ($H-K_{\rm s}$)
colour than older clusters, whose gas and dust have already been ejected.
This bimodal behaviour is also evident in the colour-magnitude diagram 
$M_K\ vs\ $($J-K_{\rm s}$), where the brightest clusters split into two sequences
separating younger from older clusters.
In addition, the reddening-free index 
$Q_{\rm d} = $($H-K_{\rm s}$)$\ -\ 0.884\ (J-H)$ has been shown to correlate
with age for the young clusters and thus provided an effective way to
differentiate the embedded clusters from the older ones.}
   {We aim to study the behaviour of these 
photometric indices for star cluster systems in the Local Group. 
In particular, we investigate the effectiveness of the $Q_{\rm d}$ index
in sorting out clusters of different ages at their early
evolutionary stages. In addition, the whole set of 
homogeneous measurements will serve as a template
for analyses of the populations belonging to distant galaxies that are
unresolved clusters or complexes. }
   {Surface photometry was carried out for 2MASS images of populous 
clusters younger than $\sim 100$\,Myr whose ages 
were available. The integrated magnitude and colours were 
measured to a limiting radius and combined to generate the photometric
diagrams. Some clusters, particularly the embedded ones,
were studied for the first time using this method.}
   {The integrated magnitudes and colours extracted from the surface photometry 
of the most populous clusters/complexes in the 
Local Group shows the expected bimodal distribution in the colour-colour and 
colour-magnitude diagrams. In particular, we confirm the index $Q_{\rm d}$ as 
a powerful
tool for distinguishing clusters younger than about 7\,Myr from
older clusters. 
}
   {}


\maketitle

\section{Introduction}

The study of 55 clusters that present strong
Br$_{\gamma}$ emission (Br$_{\gamma}-K<-0.1$) in the grand-design spiral galaxy NGC\,2997, observed with HAWK-I at VLT, has found a 
relationship between the reddening-free index $Q_{\rm d}$ and the Br$_{\gamma}$ index \citep{preben12}: 
(Br$_{\gamma}-K$)$\ = -0.48\ Q_{\rm d}\ -\ 0.17$, 
indicating that $Q_{\rm d} =$ ($H-K$)\ $-\ 0.844\ $($J-H$) represents a good age indicator
for $Q_{\rm d} > 0.1$,
which corresponds to ages $<\ $7\,Myr according to Starburst\,99 
\citep[][hereafter SB99]{leitherer99,vl05} evolutionary tracks. 
Population models generated with SB99 show nebular 
emission for the youngest clusters. 
The emission decreases with time abruptly around 7 Myr, depending on 
model parameters, as discussed in Section \ref{evol}, where errors 
to this value are also set.
\citet{preben06} previously calibrated the equivalent width of Br$_{\gamma}$, or 
equivalently,  the Br$_{\gamma}$ index (Br$_{\gamma}-K$), as an age indicator 
based on the ISAAC/VLT K-band spectra  and  SB99
evolutionary tracks. The
Br$_{\gamma}$  equivalent width is anticorrelated  with the emitting 
region age, as previously modelled for 
other H emission lines in the visible \citep{dottori81, alonsoherrero96}. 
\citet{preben12} also found that in the ($J-K$)$\ vs\ M_K$
colour-magnitude diagram (CMD) of NGC2997, the youngest clusters 
with  Q$_{\rm d}\ >\ 0.1$ concentrate around 
($J-K$)$\approx$1.8, with  significant scatter due to variable extinction. 
They form a well-defined branch that is
clearly separated from the bright older sources that concentrate around
($J-K$)$\approx$1.2. This two-branch pattern
is common to ten grand-design galaxies analysed by 
\citet{preben12} and even for 
NGC\,7424 whose cluster population is 
3 magnitudes fainter than that of NGC\,2997, as seen in 
their figures 8, 9, and B.3. 
If the extinction varied only slowly with age, 
one would not expect  a separation 
between the young and 
old branches in their diagrams. Thus, the gap 
suggests a sudden change in dust properties and/or 
distribution that lead to a rapid reduction of the extinction. 

Studies of our Galaxy, the Large Magellanic Cloud (LMC), and the Small 
Magellanic Cloud (SMC) show that the interplay between stellar 
evolution and dust properties 
in the intracluster medium is a dynamic one: stellar clusters are 
born embedded within giant molecular clouds. During their formation and 
early evolution, they are often only visible at infrared 
wavelengths, since they are heavily 
obscured by dust \citep{lada03}; moreover, the 
embedded cluster birthrate exceeds that of the visible open clusters 
by an order of magnitude \citep{lada03}.  The findings of
\citet{preben12} could help in understanding the behaviour
of the intracluster medium, which could be associated with the first supernovae
explosions and stellar winds from O-type stars, when dust 
was expelled from the cluster 
environment \citep{bastiangoodwin06, goodwinbastian06, lada03, whitworth79}.
The increase in ($H-K_{\rm s}$) for younger clusters is also a 
consequence of
the presence of hot stars because they ionise hydrogen, producing continuum
emission and the Br$_{\gamma}$ line, and also heat dust, 
both contributing to the flux in $K$.

We checked the capacity of the near-infrared (NIR) diagrams 
to differentiate clusters younger than 7\,Myr from the older ones 
in a sample of clusters from the Milky Way, the Magellanic Clouds (MCs)
and two clusters from M\,31 and M\,33, all of which have 
well-determined ages. In the next section, we present the
NIR photometry of the Local Group clusters obtained from the 2MASS images. 
In Section 3, we compare our multi-diaphragm photometry of LMC+SMC clusters
with that of \citet[][hereafter P06]{p06}. 
The age limit estimate and the NIR 
colour sensitivity on parameters/ingredients
of stellar populations evolution models are presented in Section 4. 
In Section 5, we discuss the
observational diagnostic diagrams. 
In Section 6, we show how the $Q_{\rm d}$ index can
be used as an age diagnostic tool. In Section 7, we discuss the results
and in Section 8, we present our conclusions.

\section{Near-infrared surface photometry of populous young star clusters}

The sample listed in Table \ref{LGclu} 
is based on the selection by \cite{pz10}, complemented 
with the  Milky Way embedded clusters RCW\,38, NGC\,3576 and M\,17, which 
have masses higher than 500\,M$_\odot$ \citep{lada03}, and
the LMC clusters 
NGC\,1805, 1984, 2011, 2156, 2159 and 2172, which were studied by
P06. \cite{pz10} restricted their sample to well-studied populous 
clusters with known ages; in addition to the MCs and Milky Way 
clusters,
NGC\,604 in M\,33 and vdb\,0 in M\,31  were also
analysed. All clusters have ages up to 100\,Myr.
Such a sample was considered suitable to derive the 
integrated light properties using 2MASS mosaic images built from 
relatively shallow exposures.
The coordinates, age, mass, reddening and distance 
adopted for each cluster are given in Table \ref{LGclu}.

The $JHK_{\rm s}$ images of the 42 star clusters were 
downloaded from the 2MASS database using the software Montage\footnote{Montage
  v3.3 is available in the NASA/IPAC Infrared Science Archive 
(http://irsa.ipac.caltech.edu/applications/2MASS/)}, which
assembles single images ($512\times 1024$\,pix; 1$\arcsec$/pix) into 
custom mosaics 
preserving the spatial and calibration  fidelity of the input images. 
A zeropoint keyword determined 
by the average of zeropoints provided in all single images was appended to each
mosaic file for each band.
Typically, the mosaics cover sky areas from 2$\degr \times 2\degr$ to 
12$\arcmin \times 12\arcmin$ depending on the cluster size.

The clusters' centres were determined visually in the 
images, and the surface brightness was evaluated in the annular
regions around this centre.
Fixed width annuli were employed for each cluster, but they varied 
from 2$\arcsec$ to 10$\arcsec$ depending on the 
cluster size.  
The sky background ($F_{bg} \pm \sigma_{bg}$, in DN/pix) 
was determined for each band using the whole mosaiced frame by means 
of an algorithm 
that evaluates the mode of the sky distribution and rejects pixels 
contaminated by stars.

\begin{figure}
\centering
\includegraphics[width=9cm]{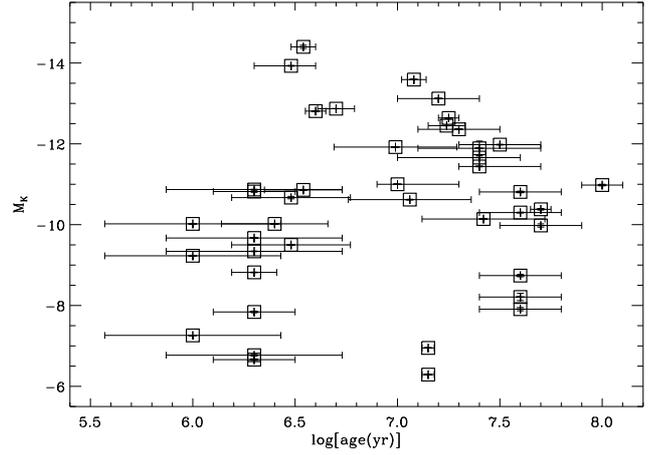}
\caption{ Absolute $K$ magnitude versus  age for 
the Local Group clusters in our sample.}
\label{Mk_age}
\end{figure}

The cluster integrated colours and their uncertainties, the 
adopted aperture radius and the calculated absolute $K$ magnitude are
given in Table \ref{LGclu}.
Fig. \ref{Mk_age} shows the distribution of the absolute $K$ magnitude 
with age for the Local Group clusters, characterising the sample in terms
of intrinsec properties.

\section{Comparison: this work vs Pessev et al. 2006}

\begin{figure}
   \centering
\includegraphics[width=9cm]{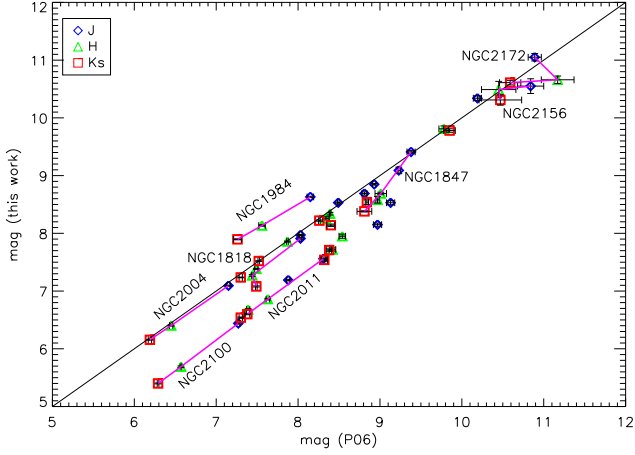}
   \caption{Comparison of our integrated near-IR magnitudes of 
MC clusters with those from P06.  $J$, $H$ and $K_{\rm s}$ magnitudes are 
plotted as different symbols and colours. The straight continuous line 
is a one-to-one relation.
Line segments connect the three magnitudes for exemplary
clusters.}
              \label{magcomp}
    \end{figure}

Fig. \ref{magcomp} shows the measured $JHK_{\rm s}$ integrated magnitudes of 
MCs' star clusters in common with those in the
P06 sample. 
In the present study, we  defined the
area covered by a cluster as limited by the radius for
which the magnitude difference from consecutive diaphragms
is approximately the same as the magnitude error. The photometry 
performed by P06, although also based on 2MASS images, followed 
a different approach: both the PSF photometry (for the resolved component)
and the surface photometry (for the unresolved component) were employed,
and probable non-members were excluded from the final set 
of aperture measurements. 
The procedure carried out by P06 to exclude probable non-member 
bright stars from the integrated photometry was based on a comparison
between the individual stars magnitudes derived from PSF photometry 
and their expected locii in a CMD considering the cluster age. Discrepancies between 
them indicated a probable non-member, which was then subtracted from
the cluster integrated light.
In our measurements, we did not exclude 
any star because our intention was to compare the clusters' integrated 
light with those obtained for clusters/complexes observed in distant
galaxies. While P06 included different error sources, even the error
introduced by background stochastic fluctuations, in the present analysis, 
only the photometric errors were considered. To quantify these errors, we 
used for each band the total noise variance as given in the 
Explanatory Supplement to the 2MASS All Sky Data 
Release\footnote{http://www.ipac.caltech.edu/2mass/releases/allsky/doc/sec6\_8a.html}, 
i.e., $\sigma_{band}=\sqrt{n\times \sigma^2_{bg}
  \times [(4\times 1.7^2) + 1.8\times 10^{-3} n]}$, where $n$ is the number of
pixels in the aperture and $\sigma_{bg}$ is the measured background noise 
(in DN/pix).
The magnitude error in a given aperture was then obtained by
$\sigma_{mag}=1.0857 \times \sigma_{band}/F_{band}$, where $F_{band}$ is
the background subtracted 
flux in DN. The calibrated integrated magnitude in each band was 
$mag_{cal}=zeropoint-2.5\log{F_{band}}$. The colour errors were 
determined by propagating the integrated
magnitude errors from the relevant bands.
Because P06 yields integrated magnitudes for several diaphragms, we 
selected the one that matches our adopted limiting radius for the
following analysis. 

\begin{figure}
   \centering
\includegraphics[width=9cm]{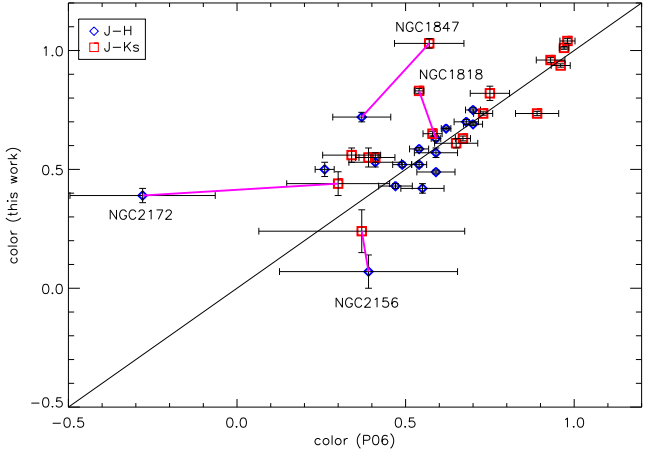}
      \caption{Comparison of our integrated near-IR colours of 
MC clusters with those from P06. ($J-H$) and ($J-K_{\rm s}$) colours are
plotted as blue diamonds and red squares, respectively. The straight 
continuous line 
is a one-to-one relationship. Line segments
connect colours of cluster outliers.}
         \label{colcomp}
   \end{figure}

The integrated magnitudes comparison is shown in Fig. \ref{magcomp}. 
There is a general agreement between both studies for all three
bands. The continuous line is a one-to-one relation.
Some clusters, taken as examples to illustrate the colours
behaviour, are indicated by line segments connecting the
three bands.  
Line segments parallel to the one-to-one relation reveal clusters 
that have identical colours in P06 and our study. Crooked or
slanted line segments yield different colours in P06 and our study. 
The colours comparison is shown in Fig. \ref{colcomp} and
indicates clusters with dissimilar colours compared to those from P06. 

\begin{figure}
   \centering
\includegraphics[width=9cm]{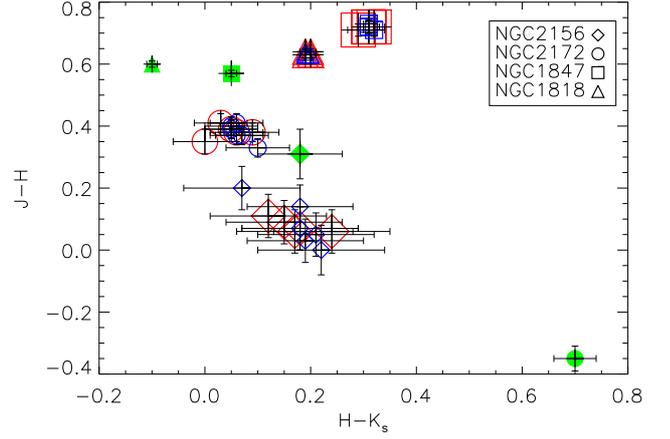}
      \caption{Colour-colour diagram of the outlier clusters indicated
in Fig. \ref{colcomp}. P06 measurements are indicated by filled green symbols.
Large open red symbols represent the effect of adopting different cluster 
centres. Small open blue symbols show the effect of different sky 
backgrounds.}
         \label{errcc}
   \end{figure}

The effect of the different adopted centres and 
background corrections
were investigated for these clusters to search for an explanation for
the discrepancies found. 
Fig. \ref{errcc} presents the results of the changing
centre and the background correction for the aforementioned clusters on 
the  ($J-H$)$\ vs\ $($H-K_{\rm s}$) diagram. The adopted cluster centre
plus four positions shifted from this centre by 10\arcsec\ towards 
the E, W, N and S, respectively, were considered, and the colours 
were determined for each of centre. The sky 
estimates were evaluated in five
different fields, defined as square areas of a typical size 2 times larger 
than the cluster 
adopted limit (see Tab. \ref{LGclu}), and the colours determined for 
the clusters maintained their original adopted centres. The colours 
derived from the original sky estimate using the whole mosaiced frame 
is also plotted, as well as the colours from P06 for these four clusters, 
where we plotted their colours for the diaphragm 
that matches our adopted cluster limit.
The effect of the moderate centre shifts and the different 
sky corrections appear not
to account for the discrepancy of the cluster colours in comparison with
the P06 measurements. Particularly, NGC\,2172 presents a huge difference
in both  colours. NGC\,2156, on the other hand, in spite the larger
photometric errors, present colours compatible with P06. NGC\,1818 and
NGC\,1847 have a disagreement in ($H-K_{\rm s}$).

\begin{figure}
   \centering
\includegraphics[width=9cm]{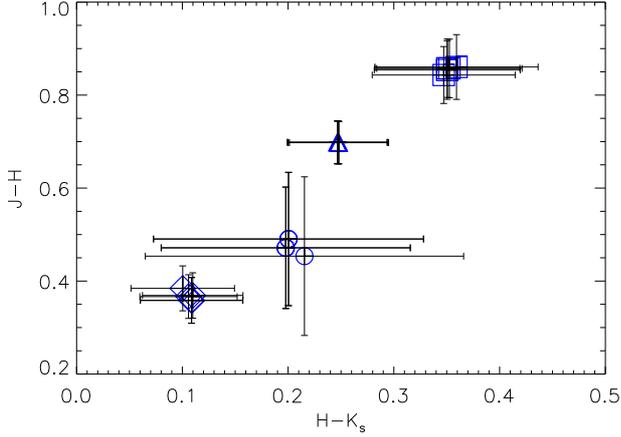}
      \caption{Colour-colour diagram of the outlier clusters indicated
in Fig. \ref{errcc} based on the integrated flux of individual stars
within the adopted cluster limit.  The symbols are the same as in
Fig. \ref{errcc}. The influence of the different centres
is shown for each cluster.}
         \label{intc}
   \end{figure}

Fig. \ref{intc} shows the ($J-H$)$\ vs\ $($H-K_{\rm s}$) diagram for 
the same clusters but with integrated colours calculated from the 2MASS
point source catalogue (PSC); i.e., the individual star flux for all stars
within the adopted cluster limit was added up. The colours resulting from 
the different centring (10\arcsec\ shifts) were also plotted.  
The different centres do not affect the cluster colours significantly. 
Because P06 did not provide a list of members and 
the star distribution in the NIR CMDs of these clusters does not define
well the expected evolutionary sequences, we did not attempt to subtract
the integrated flux from the flux of the possible non-members, 
which prevented us of making 
a direct comparison with the P06 photometry.

It is worth noticing, however, that the four clusters mentioned with colour 
discrepancies have no apparent sign of gas.
We searched for stars that could be responsible for the colour differences.
In accord to \cite{wl11} and \cite{sl09}
there are no stars with colours 
blue enough to produce a ($J-H$) colour like that of 
NGC\,2172 quoted by P06. Our integrated colours  calculated from 
surface photometry (Fig. \ref{errcc}) and individual star fluxes 
(Fig. \ref{intc}) agree, but diverge from that of P06.
As NGC\,2172 colours would fall in a rather peculiar 
position in the colour-colour 
diagram and not being fit by any model 
(see Fig. \ref{ccd_obs}), this indicates that P06
photometry may be incorrect for NGC\,2172. Because the differences between
our photometry of NGC\,1818 and NGC\,1847 and that of P06 are not as 
large as for NGC\,2172, their colour discrepancies 
may be explained by the exclusion of  bright non-member 
star(s) in P06 photometry. 
NGC\,2156 has a smaller colour discrepancy which also can be attributed to
non-member star(s) being removed in P06 photometry but not in ours photometry.

\section{Evolution of stellar populations NIR colours: model predictions}
\label{evol}

\begin{figure}
   \centering
\includegraphics[width=9cm]{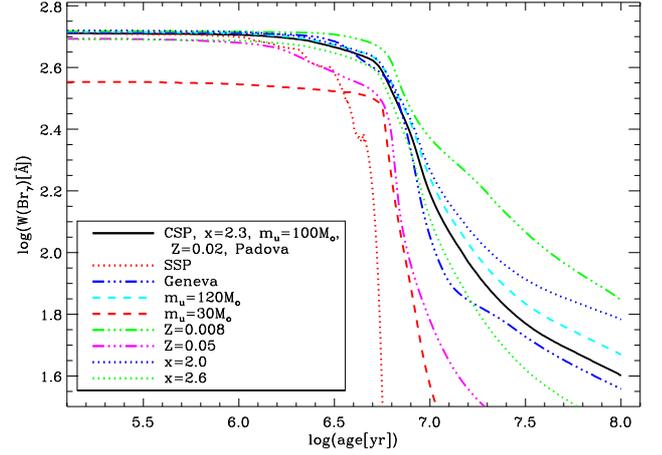}
      \caption{Equivalent width of  Br$_{\gamma}$ as a function of 
age according to SB99 models. The base model parameters/input are defined by
continuous star formation (CSP), Kroupa IMF with upper mass m$_u$=100\,M$_\odot$,
metallicity Z=0.02 and Padova tracks (continuous black line). In the following 
models, all parameters/input of the base model are kept, except for one variable
replacement:  simple stellar populations (SSP) (dotted red line);
Geneva tracks (dot-dashed blue line);
upper mass m$_u$=120\,M$_\odot$ (dashed cyan line);
upper mass m$_u$=30\,M$_\odot$ (dashed red line);
metallicity Z=0.008 (dot-dashed green line);
metallicity Z=0.05 (dot-dashed magenta line);
Kroupa IMF with upper slope x=2.0 (dotted blue line);
Kroupa IMF with upper slope x=2.6 (dotted green line).}
         \label{brg}
   \end{figure}

On theorethical grounds, it can be demonstrated that the drop of Br$_{\gamma}$
emission, associated with the evolution of hot stars that are not
capable anymore of ionising the gas in the intracluster medium, occurs at
an age of $\sim$7\,Myr. With this aim, Fig.~\ref{brg} presents the behaviour
of the Br$_{\gamma}$ equivalent width with the age according to SB99 models
(v6.0.4). 
We used as base model the one with continuous star formation 
population (CSP), standard \cite{k01} IMF (x=2.3 for masses above 0.5\,M$_\odot$
and x=1.3 for masses below 0.5\,M$_\odot$)  with upper stellar mass 
m$_u$=100\,M$_\odot$, solar metallicity (Z=0.02) and Padova \citep{bbc94,gbbc00} 
stellar evolutionary tracks. Other models for which one parameter
or ingredient is changed in the base model are represented in Fig.~\ref{brg}.
The whole set of models  
entails the effect of several variables on the age at which the  Br$_{\gamma}$
drops, according to SB99 models. We explored a range of properties that
represent a realistic approach to our problem. 

Based on the models considered in Fig.~\ref{brg}, we computed, for each 
model, the age at which the Br$_{\gamma}$ equivalent width falls by half of its 
maximum value. Then the average age and its standard deviation were 
estimated as  $7.5\pm0.9$\,Myr and considered as the age limit separating
younger populations with nebular emission from older populations without
nebular emission.

\begin{figure}
   \centering
\includegraphics[width=9cm]{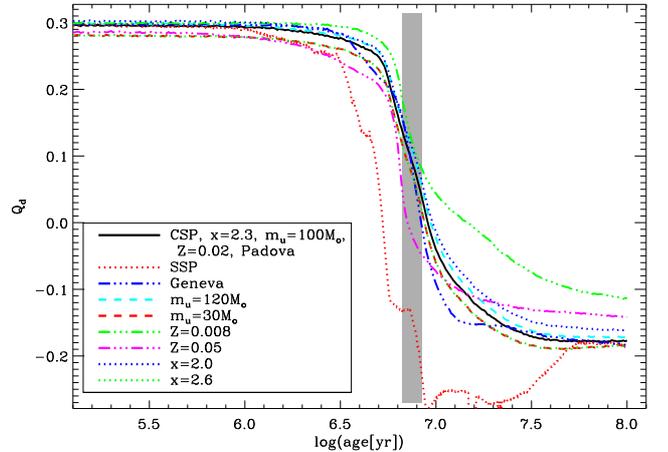}
      \caption{Reddening-free parameter as a function of age
according to SB99 models. The model parameters/input are defined as
in Fig.~\ref{brg}. The grey area represents the age (and uncertainty) estimate
for the evolutionary transition between young clusters with nebular
emission and older clusters without nebular emission.}
         \label{Qd_logt}
   \end{figure}

SB99 models also allowed us to investigate the behaviour of the reddening-free
index $Q_{\rm d}$ with age and its sensitivity to the aforementioned parameters
(Fig.~\ref{Qd_logt}). The age limit and its uncertainty as 
calculated above was used as a constraint for the range of 
$Q_{\rm d}$ representing the transition between young and old clusters. 
The higher density of models intersecting the age transition
region lies around $Q_{\rm d} = 0.1$, consistent with \cite{preben12} findings. 
However, as a cautionary note, it can be seen that in this region 
$Q_{\rm d}$ as low as  -0.15 or as high as +0.15 
are accessible to specific models. Lower values of 
$Q_{\rm d}$ favour the base model with the SSP instead of CSP as the star 
formation mode and a low upper stellar mass. Higher values of $Q_{\rm d}$ favour 
the base model (CSP, $m_u=100$\,M$_\odot$).

   \begin{figure*}
%   \centering
\includegraphics[width=9cm]{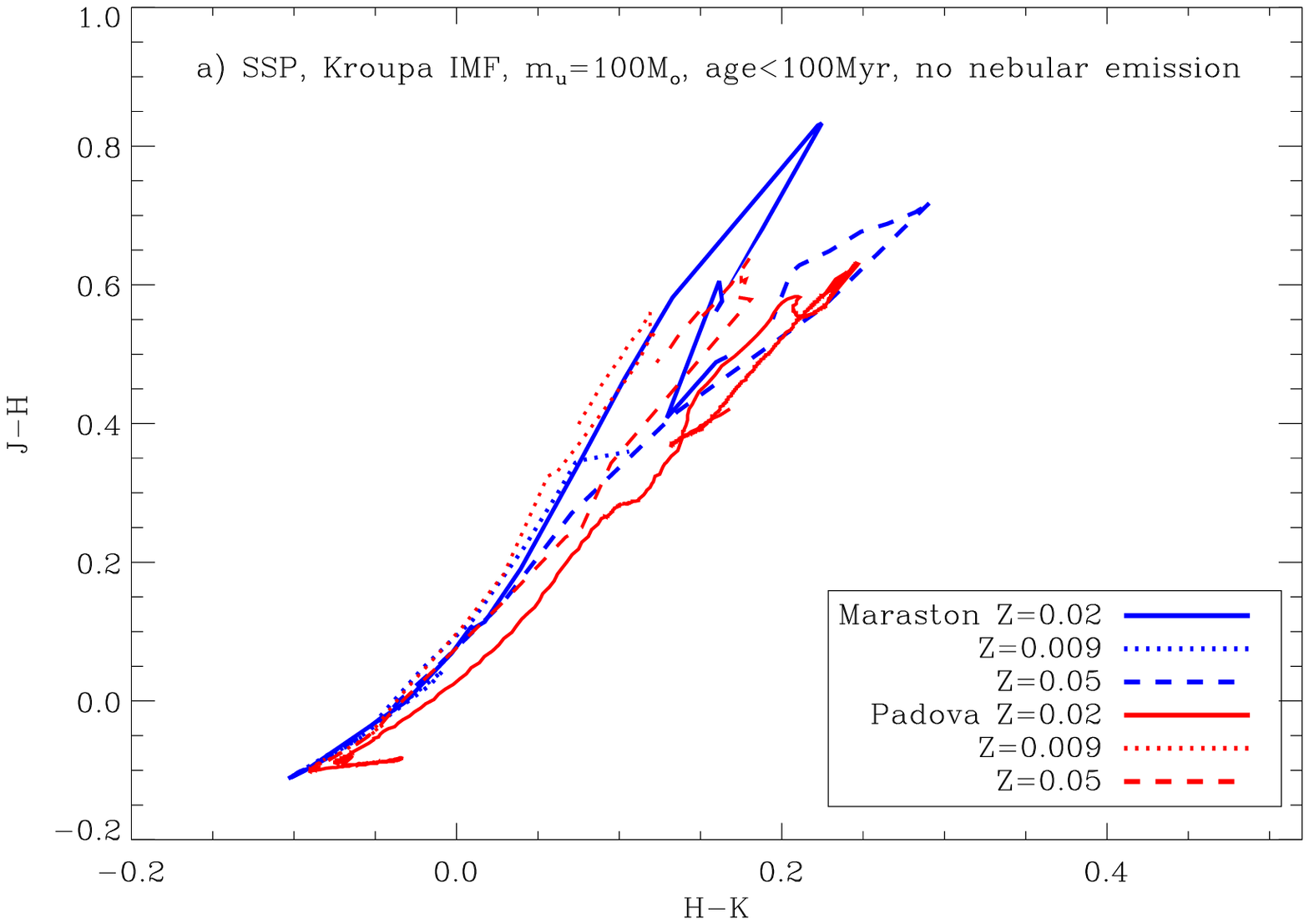}
\includegraphics[width=9cm]{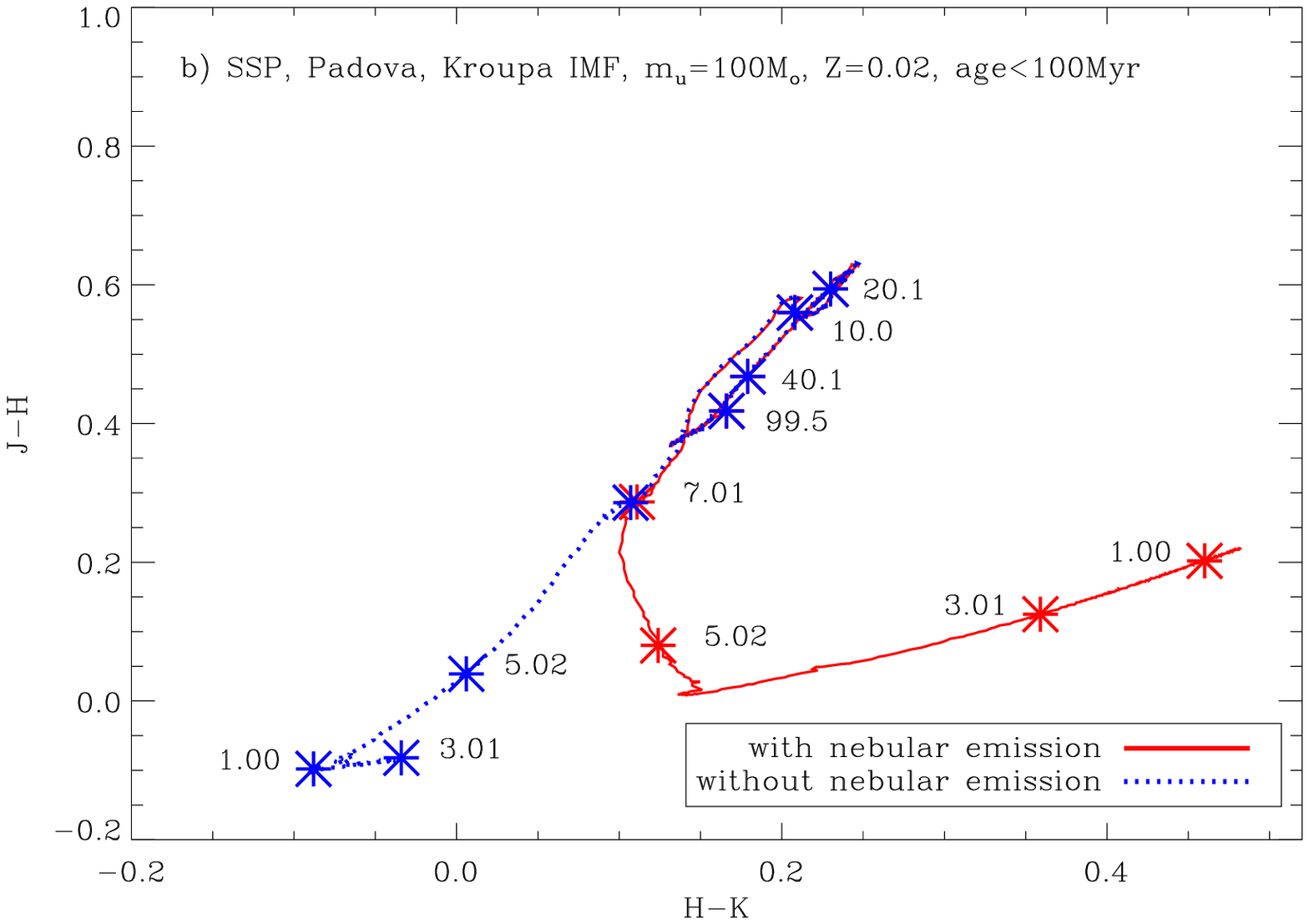}
\includegraphics[width=9cm]{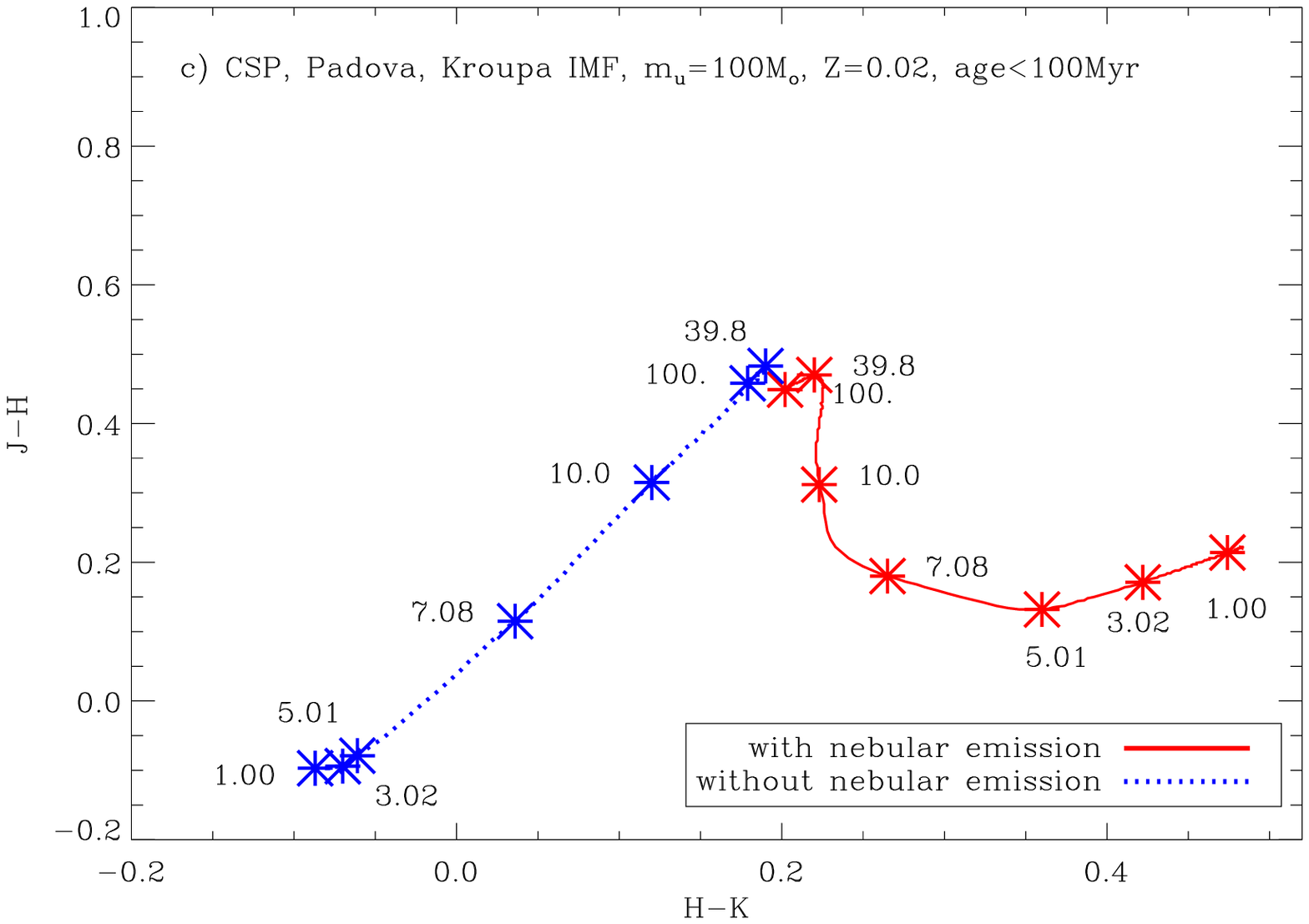}
\includegraphics[width=9cm]{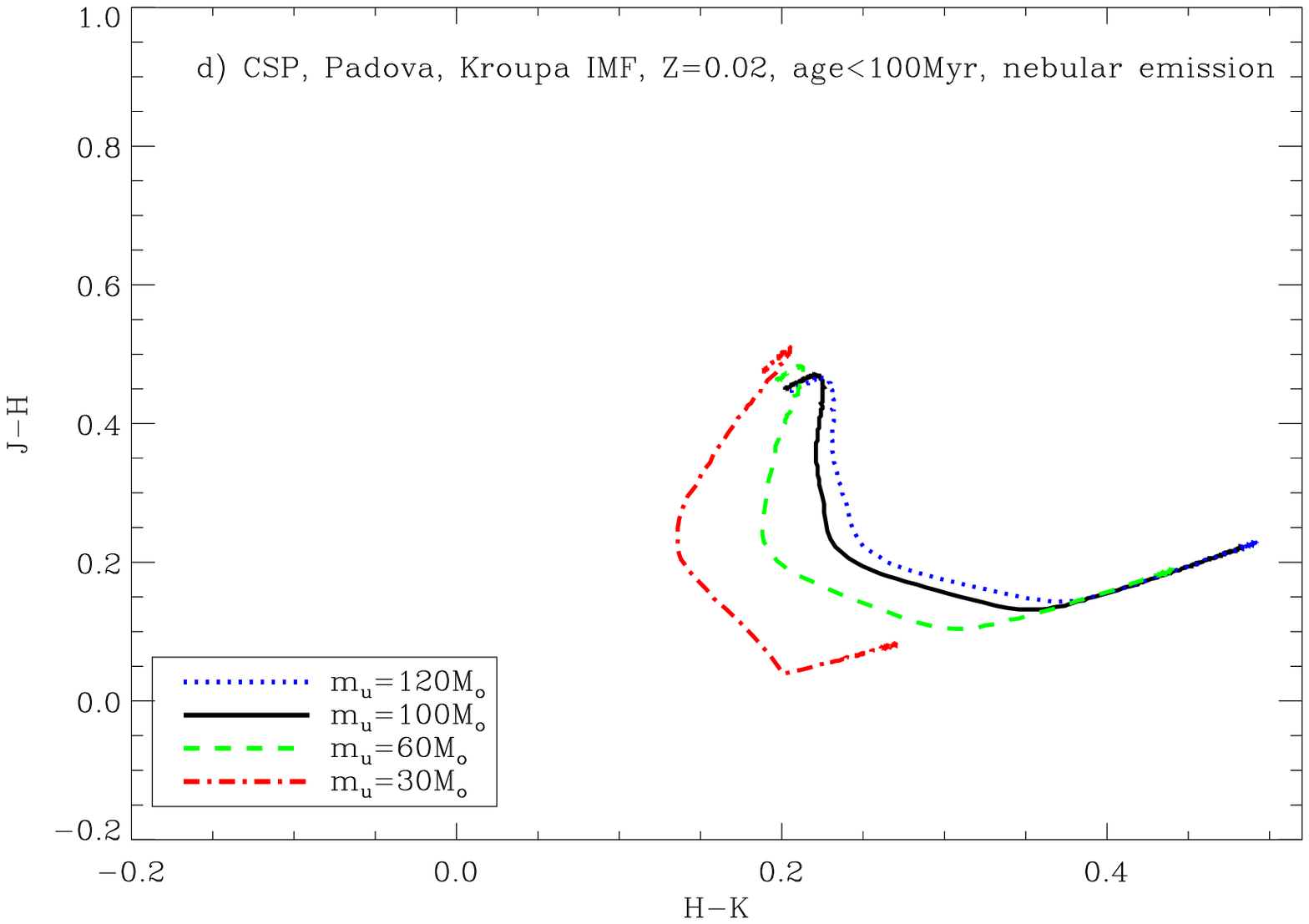}
\includegraphics[width=9cm]{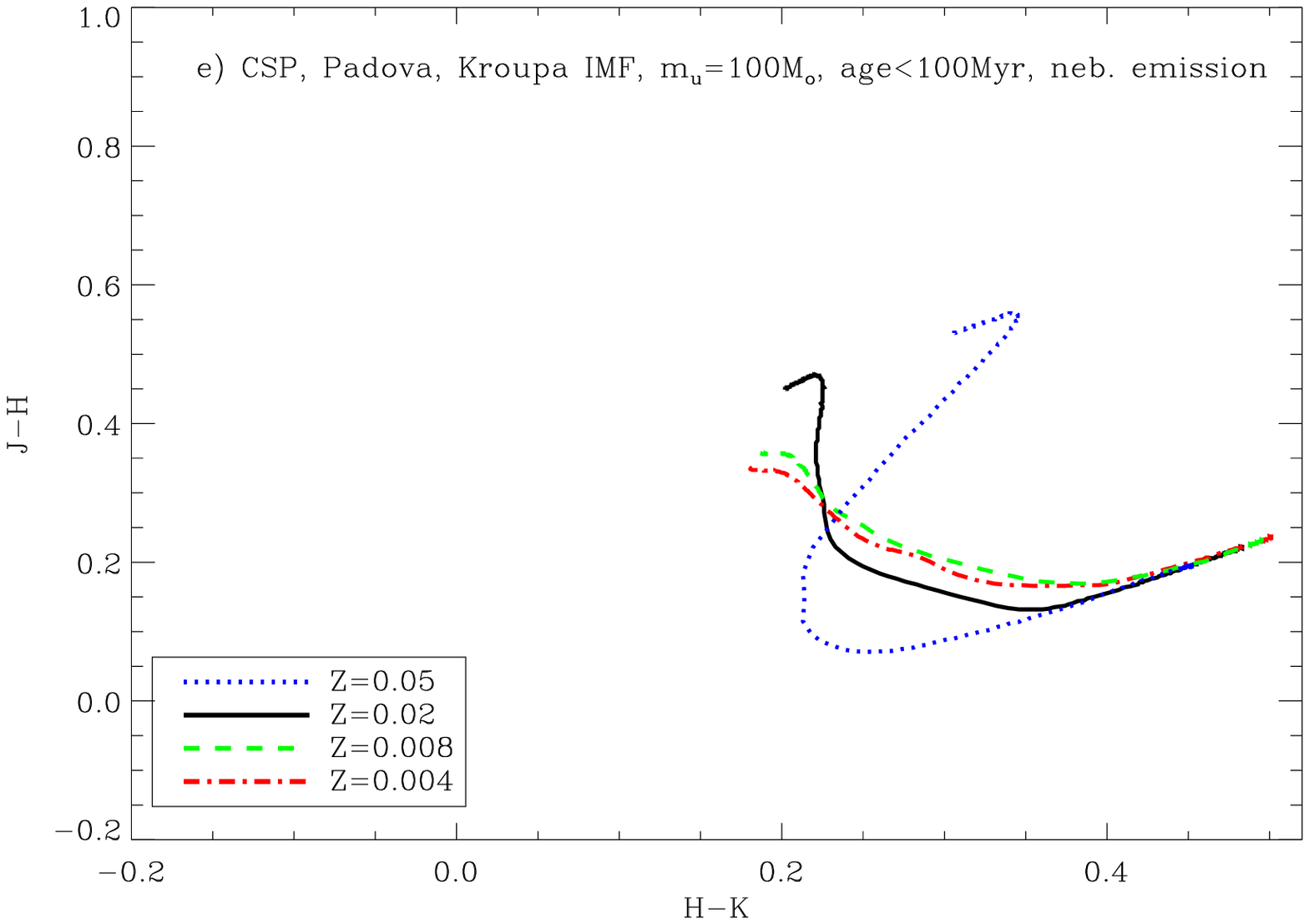}
\hskip 0.3cm \includegraphics[width=9cm]{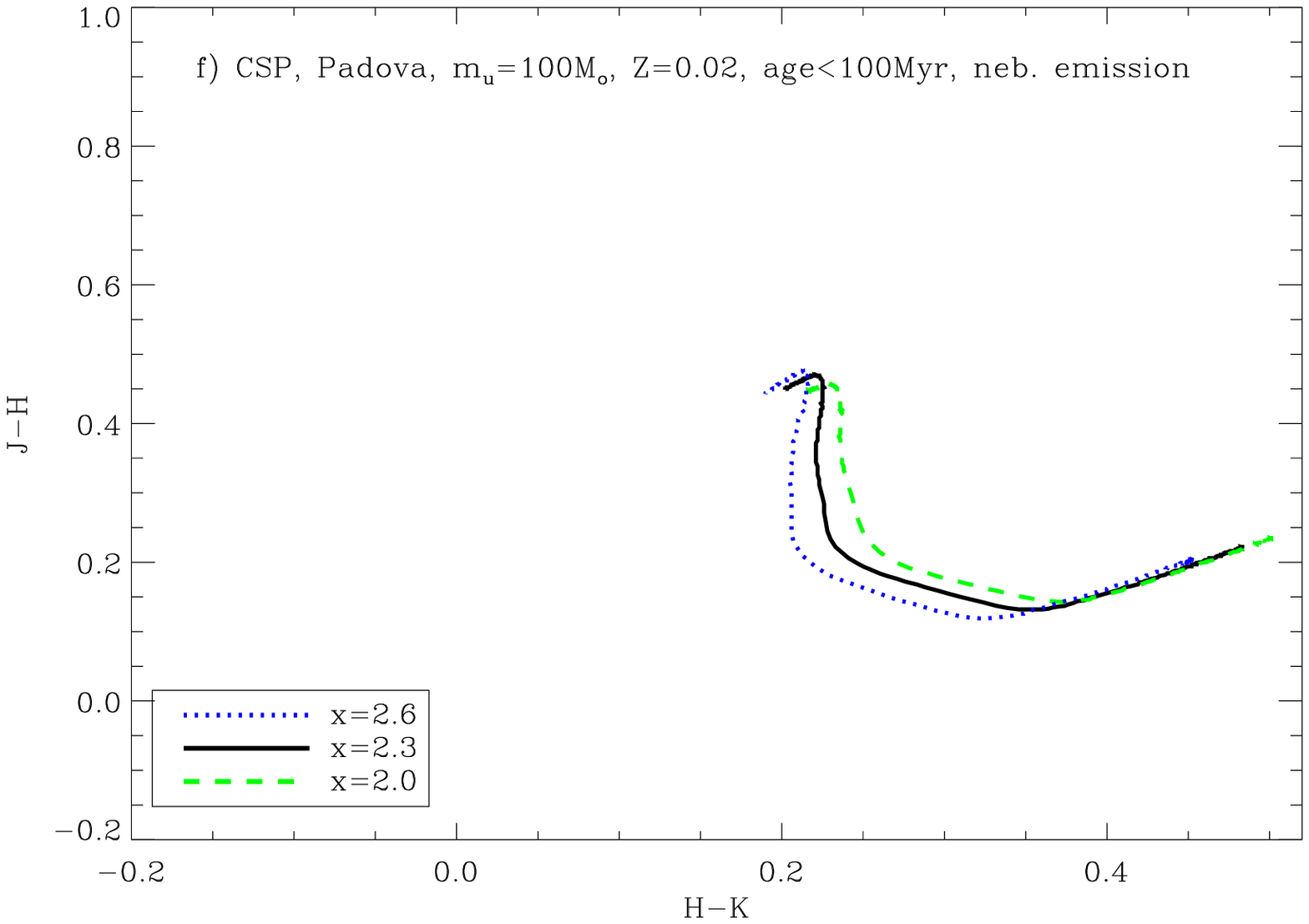}
      \caption{Colour evolution of stellar populations up to 100\,Myr 
according to different 
models and ingredients. Fixed parameters are indicated at the top while
variable ones at the bottom. (a)~Comparison between SSPs built with 
Padova tracks and SSPs from \citet{m05} for different metallicities. 
(b)~SSPs with Padova tracks including nebular emission compared to those
without nebular emission. (c)~Same as in (b) for CSPs. (d)~CSP models built
with a Kroupa IMF and variable upper stellar mass limit. (e)~~CSP models built
with different metallicities. (f)~~CSP models built
with different slopes for the IMF above 0.5\,M$_\odot$. Age (Myr) is indicated 
at selected evolutionary stages in panels (b) and (c).}
         \label{comp}
   \end{figure*}

Aiming at a comparison with the observations, the  
parameter space covered by different models  was further investigated in the 
($J-H$)$\ vs\ $($H-K_{\rm s}$) diagram (Fig. \ref{comp}).
SSP models by \citet[][hereafter M05]{m05}
are shown for  metallicities Z=0.009, 0.05 and 0.02 (solar)
and a \cite{k01} IMF and compared to SSPs built with updated Padova tracks
\citep{mgb08} 
for the same metallicities (Fig. \ref{comp}a). These models do not include
nebular emission. \cite{bekm06} showed that the M05 models would fit 
complexes in the Antennae (after reddening correction) better than 
SB99 models, which they attribute 
to the different treatment of the red supergiant phase. 
We found that Padova SSPs show a trend with metallicity opposite to those
of M05 SSPs (see the solar and above solar paths in Fig.~\ref{comp}a) 
but none of them reach extreme red colours. 
In any case, M05 models do produce redder colours for the oldest SSPs compared
to Padova SSPs, but their SSP colours are similar for populations younger than 
7\,Myr.
Fig. \ref{comp}b presents the effect of nebular emission on Padova SSP colours.
Our base SB99 model, i.e., CSPs built with Padova tracks, solar metallicity,
Kroupa's IMF with stellar mass upper limit 100\,M$_\odot$ and nebular emission
is compared to that without nebular emission in Fig. \ref{comp}c. 
Clearly the effect of nebular 
emission is to increase ($H-K_{\rm s}$) and ($J-H$) for younger populations.
We also investigated the base model colour sensitivity to the upper 
mass limit (Fig. \ref{comp}d), to the metallicity (Fig. \ref{comp}e),
and to the IMF slope above 0.5M$_\odot$ (Fig. \ref{comp}f).
($H-K_{\rm s}$) increases for the younger populations when the upper mass 
limit increases, the metallicity decreases and the IMF slope decreases.
For the range of parameters considered, the IMF slope seems to have 
the smaller effect
on the colours, but they are strongly affected by the upper mass limit.
In general, plausible variations of the parameters/ingredients investigated
are not capable of matching all Local Group cluster colours without
dust extinction (see next Sect.).

\section{Local Group clusters: the age effect on infrared colour indices}

\begin{figure}
   \centering
\includegraphics[width=9cm]{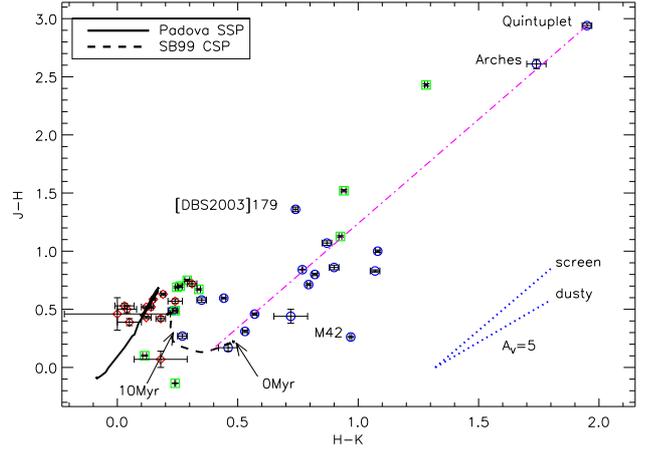}
      \caption{Sample observed colour-colour diagram with different 
symbols indicating age ranges: $t$(Myr)$<7$ (blue circle), 
$7<t$(Myr)$<20$ (green square), $20<t$(Myr)$<100$ (red diamond).
The three young clusters near the Galactic centre and M\,42 are labelled.
Continuous and dashed lines represent the colours of the Padova SSPs
and the SB99 CSP, respectively. Model ages range from very early
phases to 100\,Myr, indicated for the CSP models. 
The magenta dot-dashed line connects the observed and 
the intrinsic colours of Quintuplet. Blue dotted lines show 
the extinction yielded by $A_V=5$ for both
a standard {\it screen} reddening vector and a {\it dusty} cluster
medium.} 
         \label{ccd_obs}
   \end{figure}

\begin{figure}
\centering
\includegraphics[width=9cm]{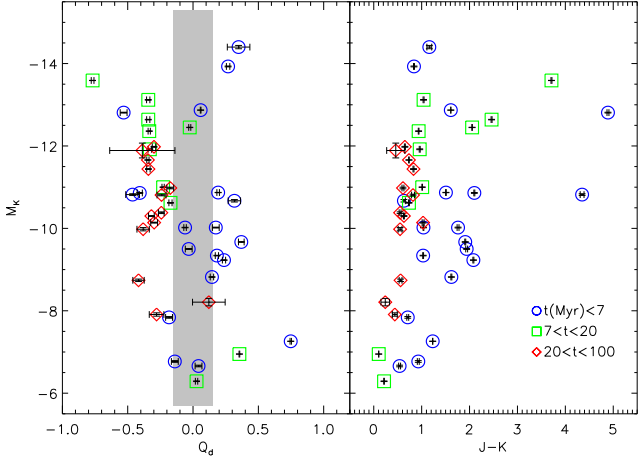}
\caption{Left: Absolute $K_{\rm s}$ band magnitudes versus the reddening-free
index $Q_{\rm d}$ for 42 clusters in the Local Group with ages up to 100\,Myr;
The grey bar indicates the range of $Q_{\rm d}$ 
values corresponding to the transition between young and old clusters,
according to Fig. ~\ref{Qd_logt}.  Right: 
Colour-magnitude diagram for the same clusters.}
\label{qvsK}
\end{figure}

Our  sample is characterised in Figs. \ref{ccd_obs} and \ref{qvsK}, 
where the measured NIR fluxes
for the 42 clusters are complemented by the ages found in the literature (see 
Table  \ref{LGclu}). The age provided come from CMD fittings of
isochrones derived from different stellar evolutionary models and the quoted 
uncertainties from the literature (large in most cases) should
account for discrepancies resulting from the use of the different 
methods of age 
determination, minimizing the effect of their heterogeneity.
We detected an error in the
age quoted by \citet{pz10} for Westerlund\,1, that  should be 
5\,Myr instead of 3.5\,Myr.

The observed colour-colour 
diagram (Fig. \ref{ccd_obs}) corroborates the explanation provided 
by \cite{preben12} for the
distribution of the clusters in grand-design spiral galaxies. 
The clusters can be separated into two main regions in the diagram.
The redder group is younger than 7\,Myr and associated with high 
extinction, and 
the bluer group is older and associated with relatively low extinction.
Figure \ref{ccd_obs} clearly shows that our sample of clusters 
younger than 7\,Myr fills regions of high extinction that are not reached
by their older counterparts.

In Fig. \ref{ccd_obs}, the colours of the Padova
single-burst stellar 
populations (SSP) with ages between 1\,Myr and 100\,Myr 
and the SB99 CSP
with ages between $\approx$\,0\,Myr and 100\,Myr are shown.
Dust and nebular emission are included in the SB99 models.  Both models 
were built using as ingredients a standard Kroupa's IMF and solar metallicity.
The different extinction models are indicated for a visual 
extinction of $A_V=5$. The standard reddening vector, the 
so-called {\it screen} model, accounts  for the foreground 
extinction \citep{indebetouw05}, while the reddening vector characterised 
by the {\it dusty} model accounts for intracluster dust, where the dust and 
stars are mixed \citep{israel98,witt92}.
The observed colours of the reddest cluster in our sample, Quintuplet,
were connected to the colours of the SB99 CSP, with an age of 4\,Myr, the 
cluster age (see Table \ref{LGclu}). Quintplets' extinction corresponds to $A_V
\approx 16$ and closely follows the {\it screen} dust model, as observed by
comparing the cluster extinction slope with that of
the extinction models. A similar argument applied to the other 
two clusters close
to the Galactic centre, Arches and [DBS2003]\,179, leads to the same
conclusion: because they are younger than 7\,Myr, linking their observed 
colours to their intrinsic colours (CSP models between 0\,Myr and 10\,Myr)
yields extinction vector slopes more akin to the {\it screen} model than
to the {\it dusty} model. 

Many clusters are compatible with both the screen and dusty models. 
The SB99 CSP models were generated with a stellar upper mass limit
of $m_u=100$\,M$_\odot$, which, in the colour-colour diagram, are suitable 
for reproducing the colours for most of the sample.
NGC\,889, NGC\,864 and NGC\,2156 are better characterised by
models with $m_u=20-30$\,M$_\odot$ and a low extinction, while M\,42
is only characterised by a young population ($0-2$\,Myr), but its colours 
are uncompatible with normal extinction. 

The distribution of the clusters in the ($J-K$)$\ vs\ M_K$ CMD, 
together with the $Q_{\rm d} \ vs \ M_K$ diagram, is shown in Fig. \ref{qvsK},
where 
the grey rectangle indicates values of $Q_{\rm d}$ corresponding to the 
transition between young and old clusters. Note that the most probable 
value is $Q_{\rm d}=0.1$ and its distribution
is not symmetric around that value.
For the ($J-H$)$\ vs\ $($H-K_{\rm s}$) diagram, Fig. \ref{qvsK} 
reveals that most of the youngest clusters (blue circles)
have a tendency to be located in specific regions of both diagrams.
In particular, 10 out of 19 clusters 
with ages less than 7\,Myr have a reddening-free index of 
$Q_{\rm d}>0.1$. This proportion does not change if we use 7.5\,Myr as
the age limit. The three very young bright clusters with 
$Q_{\rm d}<-0.3$, Arches, Quintuplet and [DBS2003]\,179, 
are objects close enough
to the Galactic centre to have their evolution influenced by
the strong tidal field, with their gas and dust already being
stripped in the early phases. They appear 
very red because of  the strong foreground extinction instead 
of their intracluster
gas/dust content. If these clusters are disregarded, then $\approx$ 60\% 
of the youngest clusters have a reddening-free index of $Q_{\rm d}>0.1$.

\begin{figure}
\centering
\includegraphics[width=9cm]{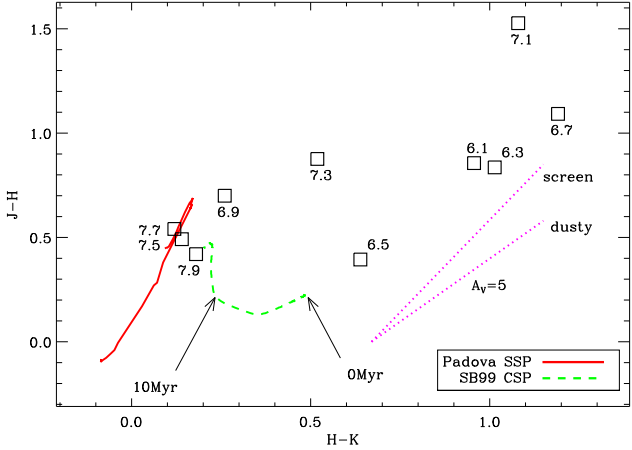}
\caption{Colour-colour diagram for age binned clusters with 
bin $\log{t}$=0.2. The age ($\log{t}$) is indicated. Padova SSP 
and SB99 CSP models are superimposed. Dust effect according to the
{\it dusty} and the {\it screen} models are represented as in 
Fig.\ref{ccd_obs}. }
\label{bin}
\end{figure}

Because less massive clusters colours may be affected by stochastic 
effects arising from small numbers of bright stars, a new colour-colour 
diagram was built (see Fig. \ref{bin}) by binning in age the cluster 
colours, computed from the integrated flux in 
$JHK$ and consequently these are flux-weigthed colours.
This procedure  guarantees that intrinsically fainter 
clusters will have low weight. To compute the final colours we did 
not weight the individual cluster colours by their signal to noise or
any other parameter.
The age bin is 0.2 in $\log{t}$.
There is a clear separation among the binned clusters
indicating two sequences in that the younger populations are redder
in ($H-K_{\rm s}$). 

\section{Age diagnostic based on $Q_{\rm d}$}

The age distribution of clusters up to 100\,Myr can be 
divided  into
clusters older ($O$) and younger ($Y$) than 7\,Myr according 
to $Q_{\rm d} < 0.1$ and $Q_{\rm d} > 0.1$, respectively.
Applying this to our observed sample of Local Group clusters, 
for $Q_{\rm d} < 0.1$,
we detect 21 true old clusters 
($O_{\rm t}$) and 6 (excluding the three clusters near the Galactic centre)
false old clusters ($O_{\rm f}$), i.e., young clusters with $Q_{\rm d} < 0.1$.
Similarly, for $Q_{\rm d} > 0.1$, we detect 10 true young clusters 
($Y_{\rm t}$) and 2
false young clusters ($Y_{\rm f}$), i.e., old clusters with $Q_{\rm d} > 0.1$.
Consequently, 78\% of the clusters having $Q_{\rm d} < 0.1$ are true old 
clusters ($O_{\rm t}/(O_{\rm t}+O_{\rm f})$), and 22\% are misclassified 
clusters younger than 7\,Myr.
In a similar way, 83\% of the clusters having $Q_{\rm d} > 0.1$ are true 
young clusters ($Y_{\rm t}/(Y_{\rm t}+Y_{\rm f})$), and 17\% are misclassified
clusters older than 7\,Myr.
The statistics also shows that, by using $Q_{\rm d} = 0.1$ as a 
discriminator, we recover $\approx$\,60\% 
($Y_{\rm t}/(Y_{\rm t}+O_{\rm f})$) from the total
sample of young clusters  and $\approx$\,90\% 
($O_{\rm t}/(O_{\rm t}+Y_{\rm f})$) from the total sample of old clusters.

Because we selected the most massive clusters and star forming 
complexes of the Local Group galaxies, the sample is, on average, representative
of the brightest (youngest) clusters in similar environments.
Consequently, beyond the qualitative agreement of our sample NIR diagrams with
those of distant galaxies \citep{preben12}, the $Q_{\rm d}$ index 
provides a reliable quantitative measure of the proportion of 
clusters/complexes younger than 7\,Myr in galaxies. 
A large fraction (83\%) of the clusters  with  $Q_{\rm d} > 0.1$ 
are genuinely 
younger than 7\,Myr and a similarly large fraction (78\%) of the clusters 
with  $Q_{\rm d} < 0.1$ are in the age range 7-100\,Myr.

\section{Discussion}

Our results validate the use of the $Q_{\rm d}$ index to differentiate
star clusters in their early stages of evolution.
The sample employed in the present study is not complete, but by selecting
the most massive clusters/complexes with available ages in the Local Group, 
we hope to have gathered an unbiased sample. One difference
from the cluster samples in \cite{preben12} is the lower 
cluster mass limit in our sample ($\approx 500$\,M$_\odot$), which 
may introduce some bias due to 
the integrated colour scattering produced by a few heavy OB stars.
Another issue is that, being lower 
metallicity galaxies, the MC clusters may 
have lower metallicity, in average, than those in the Milky Way and M31, 
making the sample not completely homogeneous on this aspect.
 
The  
$Q_{\rm d}\ vs \ M_K$ diagram (Fig. \ref{qvsK}) clearly shows a segregation of 
clusters by age and  led us to give an estimate of the proportion of clusters
younger than 7\,Myr over all clusters with $Q_{\rm d}>0.1$, which should 
be expected in galaxies outside the Local Group  if the
processes involved in cluster formation and evolution are similar.
As Fig. \ref{qvsK} shows, there is very little contamination of clusters
older than 7\,Myr among those with $Q_{\rm d}>0.1$.
It is feasible to estimate the size of the population of 
clusters younger than 7\,Myr for a distant galaxy observed in $JHK_{\rm s}$.
From the measured colours of the star clusters and the star forming regions
and the calculated  $Q_{\rm d}$ index, one can count the number of clusters
with $Q_{\rm d}>0.1$, which are $\approx 60$\% of the total number of objects
with 7\,Myr. Consequently, the total number of these objects can be 
obtained. A similar analysis for $Q_{\rm d}<0.1$ would lead to an estimate 
of the population of older clusters.
Because of their peculiar evolutionary
characteristics, the clusters near the Galactic centre were excluded 
from the above statistics, and 
the object counts in distant galaxies should exclude those clusters too
close to their centres. Nuclear clusters were also excluded in the \cite{preben12}
study of cluster systems in external galaxies.

We showed that the clusters near the Galactic centre in our sample,
i.e., Arches, Quintuplet and [DBS2003]\,179, have high extinctions associated
mainly with foreground reddening  (as opposed to intracluster dust), which nearly
follows a standard {\it screen} reddening vector (Fig. \ref{ccd_obs}).
They are clusters younger than 7\,Myr and their extreme NIR colours are
a consequence of the heavy extinction towards the Galactic centre.
The stellar population at these early ages cannot produce extremely red
colours, as demonstrated by the locus of the evolutionary models plotted 
in Fig. \ref{ccd_obs}.

In summary, if $Q_{\rm d}>0.1$, then 60\% of the clusters must be genuinely
younger than 7\,Myr, with very few older clusters showing up in this 
region (Fig. \ref{ccd_obs}). The remaining 40\% of the clusters, with an index of
$Q_{\rm d}<0.1$, can be explained by their association to extinctions consistent 
with a standard {\it screen} reddening vector. Such very young, dust
depleted clusters, as in the cases of the three clusters
near the Galactic centre in our sample, can occur as a consequence of 
extreme environmental conditions such as strong tidal 
fields. 

\section{Conclusions}
We analysed the NIR colour indexes of 42 clusters in the Local Group with ages 
determined by other 
authors to test the capacity of the reddening-free colour index
$Q_{\rm d}$ to distinguish clusters younger than 7\,Myr from the older clusters. 
The index failed to classify 6
out of the 16 clusters younger than 7\,Myr
and misclassified 2 out of the 23  clusters older than 7\,Myr. 
Because the reddening-free index verifies $Q_{\rm d}\,>\,0.1$ for 
a broad variety of galaxies, cluster masses and environments, 
we conclude that it is most likely associated more with the dust distribution 
with regard to the cluster stars than with
the metallicity or the IMF. 

The sudden change of the index is most likely linked 
to the expulsion of the dust from the cluster due
to stellar evolution, mainly SN explosions, characterising 
the transition from the mix of stars and dust inside the cluster
\citep{israel98, witt92} to the screening of the starlight by external 
dust \citep{indebetouw05}.  
It is also related to gas and dust emission, i.e. 
the presence of hot stars can excite Br$_{\gamma}$ and heat dust, 
which increases ($H-K_{\rm s}$) for younger clusters.

The $Q_{\rm d}$ index provides a way to distinguish very young 
clusters/star forming complexes up to 7\,Myr old from the older clusters 
for galaxies observed in $JHK_{\rm s}$. In addition, we showed
that the index estimates the proportion of clusters in such age ranges.

\begin{acknowledgements}
We acknowledge the anonymous referee who has helped us to 
clarify and improve significantly the paper.
We thank the GSFC Astronomy User's Library group and W. Landsman for the IDL
program used to estimate the images sky background. This research has made use
of the following databases: WEBDA, operated at the Institute for Astronomy 
of the University of Vienna; NASA/IPAC Extragalactic Database (NED) and 
 NASA/IPAC Infrared Science Archive, both operated by the Jet Propulsion 
Laboratory, California Institute of Technology, under contract with the 
National Aeronautics and Space Administration. This research made use of
Montage, funded by the NASA's Earth Science Technology Office, Computation 
Technologies Project, under Cooperative
Agreement Number NCC5-626 between NASA and the California Institute of
Technology. Montage is maintained by the NASA/IPAC Infrared Science Archive.
The english version of this  paper was revised by AJE. 
\end{acknowledgements}

\vfill\eject

\begin{landscape}
\begin{table}
\caption{Astrophysical parameters and integrated photometry of Local Group clusters.}
\label{LGclu}
\centering
\begin{tabular}{lrrrcrrrrrrcccccc}
  \hline\hline
Cluster/Galaxy &$\alpha_{2000}$ &$\delta_{2000}$ & $\log{[t(yr)]}$ & $\log[\mathcal{M}({\rm M}_\odot)]$ & $E$($B-V$) &$d$ & $R$ & $K_{\rm s}$ & $\sigma_{K_{\rm s}}$ & $M_K$&
$J-K_{\rm s}$ & $\sigma_{J-K_{\rm s}}$ & $J-H$ & $\sigma_{J-H}$ & $H-K_{\rm s}$ &
$\sigma_{H-K_{\rm s}}$ \\
& (h:m:s) &  ($\degr:\arcmin:\arcsec$) & && &(kpc) & ($\arcsec$) & &&&&&&&& \\
 \hline

NGC\,869/MW&  2:19:04 &  57:08:06 & $7.15\pm0.03$ (1)&   4.20 (18)&0.58 (20)&   2.3 (20)&   100& 5.06& 0.01&   -6.95&  0.10&  0.01& -0.14& 0.01&  0.24& 0.01\\
NGC\,884/MW&  2:20:54 &  57:08:43 & $7.15\pm0.03$ (1)&   4.10 (18)&0.56 (20)&   2.3 (20)&   140& 5.71& 0.01&   -6.29&  0.22&  0.01&  0.10& 0.01&  0.11& 0.01\\
IC\,1805/MW&  2:32:43 &  61:27:24  &  $6.30\pm0.20$ (2)&   4.20 (18)&0.82 (20)&   1.9 (20)&   600& 3.84 & 0.02 &   -7.84&  0.71 &  0.02 &  0.49 & 0.02 &  0.23 & 0.02 \\
M\,42/MW&  5:35:16 &  -5:23:15       & $6.00\pm0.43$ (3)&   3.65 (18)&0.05 (20)&   0.4 (20)&   900& 0.77& 0.01&   -7.26&  1.23&  0.01&  0.26& 0.01&  0.97& 0.01\\
NGC\,2244/MW&  6:31:55 &   4:56:36    & $6.30\pm0.20$ (2)&   3.90 (18)&0.46 (20)&   1.5 (20)&   450& 4.38 & 0.02 &   -6.66&  0.54 &  0.02 &  0.27 & 0.02 &  0.27 & 0.02 \\
RCW\,38/MW&  8:57:02 & -47:30:41    & $6.00\pm0.43$ (4)&   2.86 (19)&3.2 (21) &   1.7 (21)&   400& 3.03& 0.01&   -9.23&  2.08 &  0.01 &  1.00 & 0.01 &  1.08& 0.01\\
Westerlund\,2/MW& 10:24:02 & -57:45:29 & $6.30\pm0.11$ (2)&   4.00 (18)&0.79 (20)&   1.9 (20)&   400& 2.85& 0.01&   -8.82&  1.62 &  0.01 &  0.80 & 0.01 &  0.82 & 0.01 \\
Trumpler\,14/MW& 10:43:57 & -59:32:49 & $6.30\pm0.43$ (5)&   4.00 (18)&0.52 (20)&   2.7 (20)&   350& 2.99& 0.01&   -9.34&  1.03&  0.01&  0.46& 0.01&  0.57 & 0.01 \\
NGC\,3576/MW& 11:11:54 & -61:18:23 & $6.00\pm0.43$ (6)&   2.86 (19)&3.5 (6) &   2.6 (6)&   380& 3.27 & 0.01 &  -10.02&  1.76 &  0.02 &  0.86 & 0.02 &  0.90 & 0.02 \\
NGC\,3603/MW& 11:15:07 & -61:15:36 & $6.30\pm0.43$ (7)&   4.10 (18)&1.34 (20)&   3.6 (20)&   370& 2.38& 0.01&  -10.87&  1.51&  0.01&  0.71& 0.01&  0.79& 0.01\\
Westerlund\,1/MW& 16:47:04 & -45:50:38 & $6.70\pm0.09$ (8)&   4.50 (18)&3.0 (20) &   5.2 (20)&   160& 1.76& 0.01&  -12.87&  1.61&  0.01&  0.84& 0.01&  0.77& 0.01\\
$[$DBS2003$]$\,179/MW& 17:11:32 & -39:10:47 & $6.54\pm0.19$ (2)&  3.80 (18) &5.3 (22) &   7.9 (22)&    60& 5.47 & 0.01 &  -10.86&  2.10 &  0.01 &  1.36 & 0.02 &  0.74 & 0.01 \\
Arches/MW& 17:40:50 & -28:49:19 & $6.30\pm0.20$ (9)&  4.30 (18) &8.1 (23) &   7.6 (23)&    20& 6.40 & 0.02 &  -10.82&  4.35 &  0.04 &  2.61 & 0.04 &  1.74 & 0.04 \\
Quintuplet/MW& 17:46:15 & -28:49:39 & $6.60\pm0.05$ (9)&   4.00 (18)&8.9 (24) &   8.0 (24)&    30& 4.80& 0.01&  -12.81&  4.89 &  0.02 &  2.94 & 0.02 &  1.95 & 0.02 \\
NGC\,6611/MW& 18:18:40 & -13:46:42 & $6.48\pm0.29$ (2)&   4.40 (18)&0.78 (20)&   1.8 (20)&   530& 2.05 & 0.01 &   -9.50&  1.94 &  0.02 &  1.07 & 0.02 &  0.87 & 0.02 \\
M\,17/MW& 18:20:30 & -16:10:45 & $6.30\pm0.43$ (10)&   2.95 (19)&1.51 (20)&   1.3 (20)&   450& 1.42& 0.01&   -9.67&  1.91 &  0.01 &  0.83 & 0.01 &  1.07 & 0.02 \\
RSGC\,01/MW& 18:37:58 &  -6:53:00 & $7.08\pm0.06$ (11)&   4.50 (18)&7.9 (25) &   5.8 (25)&   130& 2.97& 0.01&  -13.59&  3.71 &  0.01 &  2.43 & 0.01 &  1.28& 0.01\\
RSGC\,02/MW& 18:39:20 &  -6:01:42 & $7.24\pm0.09$ (11)&   4.60 (18)&4.2 (26) &   5.8 (26)&   130& 2.83& 0.01&  -12.45&  2.05&  0.01&  1.13& 0.01&  0.93& 0.01\\
RSGC\,03/MW& 18:46:19 &  -3:23:16 & $7.25\pm0.05$ (12)&   4.50 (18)&4.2 (12) &   6.0 (12)&   200& 2.71& 0.01&  -12.64&  2.46 &  0.01 &  1.52 & 0.01 &  0.94& 0.01\\
Cyg\,OB\,2/MW& 20:33:15 &  41:18:50 & $6.40\pm0.26$ (2)&   4.40 (18)&2.9 (27) &   1.7 (27)&   440& 2.14& 0.01&  -10.02&  1.04&  0.01&  0.60& 0.01&  0.44& 0.01\\
NGC\,7380/MW& 22:47:15 &  58:06:34 & $6.30\pm0.43$ (2)&   3.80 (18)&0.60 (20)&   2.2 (20)&   300& 5.15 & 0.02 &   -6.77&  0.93 &  0.02 &  0.58 & 0.02 &  0.35 & 0.02 \\
NGC\,330/SMC&  0:56:19 & -72:27:49 & $7.40^{+0.20}_{-0.40}$ (13)&   4.56 (18)&0.03 (28)&  60.0 (28)&    60& 7.24& 0.01&  -11.66&  0.74&  0.01&  0.59& 0.01&  0.15& 0.01\\
NGC\,346/SMC&  0:59:05 & -72:10:40 & $6.48\pm0.29$ (14)&   5.60 (18)&0.03 (28)&  60.0 (28)&    90& 8.23 & 0.03 &  -10.67&  0.63 &  0.03 &  0.17 & 0.03 &  0.46 & 0.04 \\
NGC\,1711/LMC&  4:50:37 & -69:59:00 & $7.70\pm0.05$ (15)&   4.24 (18)&0.07 (28)&  50.0 (28)&    70& 8.14 & 0.02 &  -10.38&  0.55 &  0.02 &  0.43 & 0.01 &  0.12 & 0.02 \\
NGC\,1805/LMC&  5:02:22 & -66:06:45 & $7.00^{+0.30}_{-0.10}$ (15)&   3.52 (15)&0.07 (28)&  50.0 (28)&    50& 7.52& 0.01&  -11.00&  1.01&  0.01&  0.67& 0.01&  0.34& 0.01\\
NGC\,1818/LMC&  5:04:14 & -66:26:02 & $7.40^{+0.30}_{-0.10}$ (15)&   4.42 (18)&0.07 (28)&  50.0 (28)&    90& 7.08 & 0.01 &  -11.44&  0.83 &  0.01 &  0.63 & 0.01 &  0.19 & 0.01 \\
NGC\,1847/LMC&  5:07:08 & -68:58:23 & $7.42\pm0.30$ (15)&   4.44 (18)&0.07 (28)&  50.0 (28)&    40& 8.38 & 0.01 &  -10.14&  1.03 &  0.02 &  0.72 & 0.02 &  0.31 & 0.02 \\
NGC\,1850/LMC&  5:08:45 & -68:45:41 & $7.50\pm0.20$ (15)&   4.86 (18)&0.07 (28)&  50.0 (28)&   110& 6.54 & 0.01 &  -11.98&  0.65 &  0.01 &  0.52 & 0.01 &  0.14 & 0.01 \\
NGC\,1984/LMC&  5:27:40 & -69:08:06 & $7.06\pm0.30$ (15)&   3.38 (15)&0.07 (28)&  50.0 (28)&    40& 7.90& 0.01&  -10.62&  0.74&  0.01&  0.49& 0.01&  0.24 & 0.01 \\
\hline
\end{tabular}
\end{table}
\end{landscape}

\setcounter{table}{0}
\begin{landscape}
\begin{table}
\caption{continued.}
%\label{LGclu}
\centering
\begin{tabular}{lrrrcrrrrrrcccccc}
  \hline\hline
Cluster/Galaxy &$\alpha_{2000}$ &$\delta_{2000}$ & $\log{[t(yr)]}$ & $\log[\mathcal{M}({\rm M}_\odot)]$ & $E$($B-V$) &$d$ & $R$ & $K_{\rm s}$ & $\sigma_{K_{\rm s}}$ & $M_K$&
$J-K_{\rm s}$ & $\sigma_{J-K_{\rm s}}$ & $J-H$ & $\sigma_{J-H}$ & $H-K_{\rm s}$ &
$\sigma_{H-K_{\rm s}}$ \\
 & (h:m:s) &  ($\degr:\arcmin:\arcsec$) & && &(kpc) & ($\arcsec$) & &&&&&&&& \\
 \hline

NGC\,2004/LMC&  5:30:40 & -67:17:14 & $7.30\pm0.20$ (15)&  4.36 (18) &0.07 (28)&  50.0 (28)&   100& 6.16& 0.01&  -12.36&  0.94&  0.01&  0.69& 0.01&  0.25& 0.01\\
NGC\,2011/LMC&  5:32:19 & -67:31:18 & $6.99\pm0.30$ (15)&   3.47 (15)&0.07 (28)&  50.0 (28)&   110& 6.60 & 0.01 &  -11.92&  0.96 &  0.01 &  0.70 & 0.01 &  0.26 & 0.01 \\
NGC\,2070/LMC&  5:38:42 & -69:06:03 & $6.48^{+0.12}_{-0.18}$ (15)&   4.78 (18)&0.07 (28)&  50.0 (28)&   270& 4.59 & 0.01 &  -13.93&  0.84 &  0.01 &  0.31 & 0.01 &  0.53 & 0.01 \\
NGC\,2100/LMC&  5:42:08 & -69:12:38 & $7.20\pm0.20$ (15)&   4.36 (18)&0.07 (28)&  50.0 (28)&   170& 5.40& 0.01&  -13.12&  1.04 &  0.01 &  0.75& 0.01&  0.29 & 0.01 \\
NGC\,2136/LMC&  5:52:58 & -69:29:30 &$8.00\pm0.10$ (15)&   4.30 (18)&0.07 (28)&  50.0 (28)&   100& 7.54 & 0.02 &  -10.98&  0.61 &  0.02 &  0.42 & 0.02 &  0.18 & 0.02 \\
NGC\,2157/LMC&  5:57:35 & -69:11:46 & $7.60\pm0.20$ (15)&   4.31 (18)&0.07 (28)&  50.0 (28)&    60& 8.22 & 0.01 &  -10.30&  0.63 &  0.01 &  0.52 & 0.01 &  0.12 & 0.02 \\
NGC\,2156/LMC&  5:57:50 & -68:27:38 & $7.60\pm0.20$ (15)&   3.63 (15)&0.07 (28)&  50.0 (28)&    60&10.31 & 0.09 &   -8.21&  0.24 &  0.09 &  0.07 & 0.07 &  0.18 & 0.11 \\
NGC\,2159/LMC&  5:58:03 & -68:37:22 & $7.60\pm0.20$ (15)&   3.65 (15)&0.07 (28)&  50.0 (28)&    40& 9.78 & 0.03 &   -8.74&  0.56 &  0.03 &  0.53 & 0.02 &  0.03 & 0.04 \\
NGC\,2164/LMC&  5:58:55 & -68:30:57 & $7.70\pm0.20$ (15)&   4.18 (18)&0.07 (28)&  50.0 (28)&    90& 8.54 & 0.04 &   -9.98&  0.55 &  0.04 &  0.50 & 0.03 &  0.04 & 0.04 \\
NGC\,2172/LMC&  6:00:05 & -68:38:13 & $7.60\pm0.20$ (15)&   3.52 (15)&0.07 (28)&  50.0 (28)&    30&10.61 & 0.04 &   -7.91&  0.44 &  0.05 &  0.39 & 0.03 &  0.05 & 0.05 \\
NGC\,2214/LMC&  6:12:58 & -68:15:36 & $7.60\pm0.20$ (15)&   4.03 (18)&0.07 (28)&  50.0 (28)&   110& 7.71 & 0.02 &  -10.81&  0.82 &  0.03 &  0.57 & 0.02 &  0.24 & 0.03 \\
VdB\,0/M\,31&  0:40:29 &  40:36:15 & $7.40\pm0.30$ (16)&   4.85 (18)&0.05 (28)& 783.0 (28)&    20&12.65 & 0.18 &  -11.89&  0.46 &  0.19 &  0.46 & 0.14 &  0.00 & 0.22 \\
NGC\,604/M\,33&  1:34:33 &  30:47:00 & $6.54\pm0.06$ (17)&   5.00 (18)&0.04 (28)& 883.0 (28)&    40&10.34 & 0.04 &  -14.40&  1.16 &  0.05 &  0.44 & 0.06 &  0.72 & 0.07 \\
\hline
\end{tabular}
\tablefoot{Literature sources for age, photometric mass, reddening and 
distance are indicated within parentheses.}
\tablebib{(1)~\citet{chi10};
(2) \citet{p09}; (3) \citet{hh98}; (4) \citet{wsba06};
(5) \citet{aavl07}; (6) \citet{fbdc02}; (7) \citet{hem08};
(8) \citet{mt07}; (9) \citet{fmm99}; (10) \citet{pcb09};
(11) \citet{f08}; (12) \citet{cnd09}; (13) \citet{mg03b}; (14) \citet{ssn08};
(15) \citet{mg03a}; (16) \citet{pbb09}; (17) \citet{m01};
(18) \citet{pz10}; (19) \citet{lada03}; (20) WEBDA
  (http://www.univie.ac.at/webda/); (21) \citet{sbw99};
(22) \citet{bih08}; (23) \citet{mhp08}; (24) \citet{lho12};
(25) \citet{fmr06}; (26) \citet{dfk07}; (27) \citet{k00};
(28) NED/NASA (http://ned.ipac.caltech.edu/).
}
\end{table}
\end{landscape}


\begin{thebibliography}{}
\bibliographystyle{aa} 
 
  \bibitem[Alonso-Herrero et al.(1996)]{alonsoherrero96} Alonso-Herrero, A., Aragon-Salamanca, A., Zamorano, J.
   \& Rego, M. 1996, MNRAS, 278, 417 

 \bibitem[Ascenso et al.(2007)]{aavl07}
Ascenso, J., Alves, J., Vicente, S., Lago, M. T. V. T. 2007, A\&A, 476, 199


 \bibitem[Bastian et al.(2006)]{bekm06}
Bastian, N., Emsellem, E., Kissler-Patig, M., Maraston, C. 2006, A\&A 445, 471


  \bibitem[Bastian \& Goodwin(2006)]{bastiangoodwin06} Bastian, N., Goodwin, S.P. 2006, MNRAS, 367, L9

 \bibitem[Bertelli et al.(1994)]{bbc94}
Bertelli, G., Bressan, A., Chiosi, C., Fagotto, F., Nasi, E. 1994, 
A\&AS 106, 275 

  \bibitem[Borissova et al.(2008)]{bih08}
Borissova, J., Ivanov, V. D., Hanson, M. M., et al. 2008, A\&A, 488, 151

  \bibitem[Chandar et al.(2010)]{cwf10}
Chandar, R., Whitmore, B.C., Fall, S.M. 2010, ApJ, 713, 1343

  \bibitem[Clark et al.(2009)]{cnd09}
Clark, J. S., Negueruela, I., Davies, B., et al. 2009, A\&A, 498, 109

  \bibitem[Currie et al.(2010)]{chi10}
Currie, T., Hernandez, J., Irwin, J. 2010, ApJS, 186, 191

  \bibitem[Davies et al.(2007)]{dfk07}
Davies, B., Figer D. F., Kudritzki, R.P., et al.  2007, ApJ, 671, 781

  \bibitem[Dottori(1981)]{dottori81} Dottori, H.A. 1981, Ap\&SS, 80, 267

  \bibitem[Figer(2008)]{f08}
Figer, D. F. 2008, in Massive Stars as Cosmic Engines, ed. F. Bresolin, P. A. Crowther, J. Puls, Proc. IAU Symp. 250, p. 247

  \bibitem[Figer et al.(2006)]{fmr06}
Figer, D.F., MacKenty, J. W., Robberto, M., et al. 2006, ApJ, 643, 1166

 \bibitem[Figer, McLean \& Morris(1999)]{fmm99}
Figer, D. F., McLean, I. S., Morris, M. 1999, ApJ, 514, 202

  \bibitem[Figueredo et al.(2002)]{fbdc02} 
Figueredo, E., Blum, R.D., Damineli, A., Conti, P.S. 2002, AJ, 124, 2739

 \bibitem[Girardi et al.(2000)]{gbbc00}
Girardi, L., Bressan, A., Bertelli, G., Chiosi, C. 2000, A\&AS, 141, 371


  \bibitem[Goodwin \& Bastian(2006)]{goodwinbastian06} Goodwin, S.P., Bastian, N. 2006, MNRAS, 373,752 

   \bibitem[Grosb{\o}l, Dottori \& Gredel(2006)]{preben06} Grosb{\o}l, P., Dottori, H., Gredel, R. 2006, A\&A, 453, L25

  \bibitem[Grosb{\o}l \& Dottori(2012)]{preben12} Grosb{\o}l, P, Dottori, H. 2012, A\&A, 542, 39

 \bibitem[Harayama, Eisenhauer \& Martins(2008)]{hem08}
Harayama, Y., Eisenhauer, F., Martins, F. 2008, ApJ, 675, 1319

 \bibitem[Hillenbrand \& Hartmann(1998)]{hh98}
Hillenbrand, L. A., Hartmann, L.W. 1998, ApJ, 492, 540

  \bibitem[Indebetouw et al.(2005)]{indebetouw05} Indebetouw, R., Mathis, J.S., Babler, B.L., et al. 2005, ApJ 619, 931

   \bibitem[Israel et al.(1998)]{israel98} Israel, F.P., van den Werf, P.P., Hawarden, T.G. \& Aspin, C. 1998, A\&A 336, 433

   \bibitem[Kn\"odlseder(2000)]{k00} Kn\"odlseder J. 2000, A\&A, 360, 539

 \bibitem[Kroupa(2001)]{k01}
Kroupa P. 2001, MNRAS, 322, 231
 
   \bibitem[Lada \& Lada(2003)]{lada03} Lada, C.J., Lada, E.A. 2003, ARA\&A, 41, 57

   \bibitem[Leitherer(1999)]{leitherer99} Leitherer, C., Schaerer, D., Goldader, J.D., et al. 1999, ApJS, 123, 3

   \bibitem[Liermann et al.(2012)]{lho12}
 Liermann, A., Hamann, W. R., Oskinova, L. M. 2012, A\&A, 540A, 14 

   \bibitem[Mackey \& Gilmore(2003a)]{mg03a}
Mackey, A. D., Gilmore, G. F. 2003, MNRAS, 338, 85

  \bibitem[Mackey \& Gilmore(2003b)]{mg03b}
Mackey, A. D., Gilmore, G. F. 2003, MNRAS, 338, 120

 \bibitem[Ma\'iz-Apell\'aniz(2001)]{m01}
Ma\'iz-Apell\'aniz, J. 2001, ApJ, 563, 151

   \bibitem[Maraston(2005)]{m05}
Maraston, C. 2005, MNRAS, 362, 799

   \bibitem[Marigo et al.(2008)]{mgb08}
Marigo, P., Girardi, L., Bressan, A. 2008, A\&A, 482, 883 

   \bibitem[Martins et al.(2008)]{mhp08}
Martins, F., Hillier, D.J., Paumard, T., Eisenhauser, F., Ott, T., Genzel,
R. 2008, A\&A, 478, 219

  \bibitem[Mengel \& Tacconi-Garman(2007)]{mt07}
Mengel, S., Tacconi-Garman, L. E. 2007, A\&A, 466, 151

 \bibitem[Perina et al.(2009)]{pbb09}
Perina, S., Barmby, P., Beasley, M. A., et al. 2009, A\&A, 494, 933

   \bibitem[Pessev et al.(2006)]{p06} Pessev, P. M., Goudfrooij, P., Puzia, T. H. \& Chandar
    R. 2006, AJ, 132, 781

 \bibitem[Pfalzner(2009)]{p09}
Pfalzner S. 2009, A\&A, 498, L37

   \bibitem[Portegies Zwart et al.(2010)]{pz10} Portegies Zwart, S. F., McMillan, S. L. W. \& Gieles,
     M.  2010, ARA\&A, 48, 431

   \bibitem[Povich et al.(2009)]{pcb09}
 Povich, M.S., Churchwell, E., Bieging, J.H., et al. 2009, ApJ, 696, 1278

 \bibitem[Sabbi, Sirianni \& Nota(2008)]{ssn08}
Sabbi, E., Sirianni M., Nota A. 2008, AJ, 135, 173 

 \bibitem[Salpeter(1955)]{s55}
Salpeter, E. 1955, ApJ, 121, 161

   \bibitem[Smith et al.(1999)]{sbw99}
Smith, C.H., Bourke, T.L., Wright, C.M., et al. 1999, MNRAS, 303, 367
 
   \bibitem[Straizys \& Lazauskait\.e(2009)]{sl09}
Straizys, V., Lazauskait\.e, R. 2009, Baltic Astronomy, 18, 19 

 \bibitem[Vazquez \& Leitherer(2005)]{vl05}
Vazquez, G.A., Leitherer, C. 2005, ApJ, 621, 695  

   \bibitem[Whitworth(1979)]{whitworth79}
Whitworth, A. 1979, MNRAS, 186, 59

   \bibitem[Witt, Thronson \& Capuano(1992)]{witt92} Witt, A.N., Thronson, H.A. \& Capuano, Jr., J.M. 1992, ApJ, 393, 611

   \bibitem[Wolk et al.(2006)]{wsba06}
Wolk, S. J., Spitzbart, B. D., Bourke, T. L., Alves, J. 2006, ApJ, 132, 110

   \bibitem[Worthey \& Lee(2011)]{wl11}
Worthey, G., Lee H.-C. 2011, ApJS, 193, 1

\end{thebibliography}
\end{document}